\documentclass[onecolumn,12pt]{elsarticle}
\usepackage{graphicx}
\usepackage{verbatim}
\usepackage{natbib}
\usepackage{amssymb}
\usepackage{amsmath}
\usepackage{times}
\usepackage{pstricks}
\usepackage{epsfig}
 \usepackage{epsf}
\usepackage{enumerate}
\def\arcsec{\hbox{$^{\prime\prime}$}}

\begin{document}

\begin{frontmatter}

\title{Space-quality data from balloon-borne telescopes:  \\ the \emph{High Altitude Lensing Observatory} (HALO)}

\author[jpl,cit]{Jason Rhodes}
\author[jpl,cit]{Benjamin Dobke}
\author[jpl]{Jeffrey Booth}
\author[dur3,ifa]{Richard Massey}
\author[jpl]{Kurt Liewer}
\author[cit]{Roger Smith}
\author[eth]{Adam Amara}
\author[jpl]{Jack Aldrich}
\author[eth]{Joel Berge}
\author[ukatc]{Naidu Bezawada}
\author[jpl]{Paul Brugarolas}
\author[dur]{Paul Clark}
\author[dur]{Cornelis M. Dubbeldam}
\author[cit]{Richard Ellis}
\author[dur2]{Carlos Frenk}
\author[ukatc]{Angus Gallie}
\author[ifa]{Alan Heavens}
\author[ukatc]{David Henry}
\author[oamp]{Eric Jullo}
\author[ifa]{Thomas Kitching}
\author[wff]{James Lanzi}
\author[eth]{Simon Lilly}
\author[ukatc]{David Lunney}
\author[naoj]{Satoshi Miyazaki}
\author[e2v]{David Morris}
\author[jpl]{Christopher Paine}
\author[ifa]{John Peacock}
\author[cit]{Sergio Pellegrino}
\author[e2v]{Roger Pittock}
\author[e2v]{Peter Pool}
\author[eth]{Alexandre Refregier}
\author[jpl]{Michael Seiffert}
\author[dur,dur3]{Ray Sharples}
\author[cit]{Alexandra Smith}
\author[wff]{David Stuchlik}
\author[ifa]{Andy Taylor}
\author[ipac]{Harry Teplitz}
\author[kicp]{R. Ali Vanderveld}
\author[jpl]{James Wu}

\address[jpl]{Jet Propulsion Laboratory, California Institute of Technology 4800 Oak Grove Drive, Pasadena, CA 91109, United States}
\address[cit]{California Institute of Technology, 1201 E. California Blvd., Pasadena, CA 91125, United States}
\address[dur3]{Department of Physics, University of Durham, South Road, Durham DH1 3LE, United Kingdom}
\address[ifa]{Institute for Astronomy, University of Edinburgh, Royal Observatory  Blackford Hill, Edinburgh, EH9 3HJ, United Kingdom}
\address[eth]{ETH Zurich, Institute of Astronomy, Physics Department, Wolfgang-Pauli-Strasse 27, CH-8093 Zurich, Switzerland}
\address[ukatc]{UK Astronomy Technology Centre, Royal Observatory, Blackford Hill, Edinburgh EH9 3HJ, United Kingdom}
\address[dur]{Centre for Advanced Instrumentation, University of Durham, Joseph Swan Road, Netpark, Sedgefield TS21 3FB, United Kingdom}
\address[dur2]{Institute for Computational Cosmology, University of Durham, South Road, Durham DH1 3LE, United Kingdom}
\address[oamp]{Laboratoire d'Astrophysique de Marseille, Universit\'e de Provence, CNRS, 13388 Marseille CEDEX 13, France}
\address[wff]{NASA Wallops Flight Facility, Wallops Island, VA 23337, United States}
\address[naoj]{National Astronomical Observatory of Japan, Mitaka, Tokyo, 181-8588, Japan}
\address[e2v]{e2v, 106 Waterhouse Lane, Chelmsford, Essex, CM1 2QU, United Kingdom}
\address[ipac]{Infrared Processing and Analysis Center (IPAC), California Institute of Technology, Mail Code 100-22, 770 South Wilson Avenue, Pasadena, CA 91125, United States}
\address[kicp]{Kavli Institute for Cosmological Physics, Enrico Fermi Institute, University of Chicago, 933 East 56th Street, Chicago, IL 60637, United States}\vspace{1in}


\begin{abstract}
We present a method for attaining sub-arcsecond pointing stability during sub-orbital balloon flights, as designed for in the \emph{High Altitude Lensing Observatory} (HALO) concept. The pointing method presented here has the potential to perform near-space quality optical astronomical imaging at $\sim$1--2\% of the cost of space-based missions. We also discuss an architecture that can achieve sufficient thermo-mechanical stability to match
the pointing stability. This concept is motivated by advances in the development and testing of Ultra Long Duration Balloon (ULDB) flights which promise to allow observation campaigns lasting more than three months.   The design incorporates a multi-stage pointing architecture comprising:  a gondola coarse azimuth control system, a multi-axis nested gimbal frame structure with arcsecond stability, a telescope de-rotator to eliminate field rotation, and a fine guidance stage consisting of both a telescope mounted angular rate sensor and guide CCDs in the focal plane to drive a fast-steering mirror.   We discuss the results of pointing tests together with a preliminary thermo-mechanical analysis required for sub-arcsecond pointing at high altitude.  Possible future applications in the areas of wide-field surveys and exoplanet searches are also discussed.
\end{abstract}


\end{frontmatter}

\section{Introduction}
\label{sect:intro}

The advent of space-based
observations has transformed the fields of Earth science, planetary science and astronomy.  The significantly enhanced image clarity, unique vantage point, and 24-hour observation capabilities offered in Earth orbit  resulted in an  understanding  of our planet and our place in the Universe to unprecedented levels.
However, these advances come at a significant cost.  For any space-based instrument or experiment to be feasible, even the most cost-effective design and implementation requires substantial financial resources and launch facilities often only found at the national agency level.

For a number of years, a far cheaper alternative has been to fly experiments and instrumentation aboard sub-orbital balloon flights.  In rising above the majority of the Earth's atmosphere, many of the advantages of space-based missions can be achieved at a small fraction of the cost of a fully-fledged orbiting space platform.  Despite this apparent low-cost advantage, two key issues have limited sub-orbital balloon flights for the purpose of precision optical astronomy:  flight duration and pointing stability.  The former results in a significant reduction in the total {\it{quantity}} of scientific data that can be recorded, whereas the latter results in a significant reduction in overall {\it{quality}} of those data, when compared to the space-based equivalent.  However, with the development and testing of Ultra Long Duration Balloon (ULDB) flights underway, the observation campaigns of the near future could be several months  (up to $\sim100$ days) in duration for each launch, greatly increasing the data-gathering capabilities.  Assuming a successful recovery of the payload, HALO could conduct an $\sim100$ day observing run (flight) each year, with of order 1000 square degrees observed in each flight. To address the issue of data quality, this paper introduces the design approach for a high-precision pointing stability architecture that was developed for the \emph{High Altitude Lensing Observatory} (HALO) concept.

The HALO concept is a \emph{weak lensing} balloon-borne mission. Weak lensing is the phenomenon by which the deflection of light from faint background galaxies by foreground dark matter structures. HALO has the ultimate goal of surveying several thousand square degrees of the night sky to high-precision in order to constrain the nature of dark energy (the name given to the mysterious force or property of space-time that is causing the expansion of the Universe to accelerate).  While there are a number of ways to study dark energy, weak lensing is potentially the most effective at constraining dark energy and possible modifications to General Relativity, provided that systematic effects are understood and mitigated (\cite{albr}; \cite{peac}).
 Weak lensing distorts the apparent shapes of background galaxies by a few percent, informing us of the intervening dark matter, while the time-dependent growth of dark matter clustering probes the nature of dark energy.
 Weak lensing as probe of dark energy is one of the cleanest probes as the lensing signal depends only on the underlying total mass distribution and there are no significant astrophysical sources of uncertainty or contamination.  Rather, weak lensing requires precision in the measurement of galaxy shapes and photometric redshifts.
 Given the small (percent level) effects of weak lensing on  galaxy shapes, a small and well-modeled  Point Spread Function (PSF) is absolutely crucial for successful measurement.
 A balloon-borne experiment like HALO, with a $\sim1$m class mirror and flying above more than 99.9\% of the atmosphere can achieve a PSF size of about $0.2\arcsec$, compared to a PSF size of about $0.7\arcsec$ from the ground.
 The requirement on PSF knowledge in practice can be met by placing a stringent requirement on PSF stability,  which has been notoriously difficult to maintain in ground-based experiments(see, e.g. \cite{fu};\cite{tere};\cite{heym}).
 The requirements on PSF size and stability drive weak lensing mission design to require observations made above the deleterious effects of the atmosphere for maximum control of shape measurement systematics.  This control of systematics will be enabled by the pointing developments and thermo-mechanical architecture outlined in this paper and will allow HALO to achieve  imaging resolution approaching that of the HST-COSMOS survey \citep{scov} with a survey area an appreciable fraction of the Sloan Digital Sky Survey (SDSS; \cite{york}). This architecture also has the potential to advance data collection in any field of astronomy requiring high pointing stability.  Thus, the goal of this paper is to show that one aspect of a space mission (PSF
 size and stability) can be approached with a balloon-borne system.

This paper has the following layout: In section 2, we provide a short background to sub-orbital balloon flights and some previous applications.  In section 3, we outline HALO requirements and describe the specifics of a new technical approach proposed to attain the required  sub-arcsecond pointing stability and thermo-mechanical stability, along with preliminary results of in-lab testing and modeling.  Section 4 discusses the future applications and the possible science-impact of such a system.  We give a brief summary and discussion in section 5.

\section{Background to sub-orbital balloon flights}
\label{sect:back}

\subsection{Early and continuing development}
\label{sect:develop}

Scientific balloon flights have been around in one form or another since the late seventeenth century and, before the advent of orbital satellites, were the best remote method for weather data recording at altitude.  While the basics of scientific balloon experiments have not changed, balloon capabilities have increased and their dependability has improved greatly.  They have the advantage of being able to be launched from locations worldwide to support scientific studies, the development and readiness timescales are relatively short, and crucially, the payload/instrumentation are very often reusable and upgradeable.

Present-day standard NASA scientific balloons are constructed of polyethylene film of $0.002$ centimeters in thickness and filled with helium gas.
There are currently three types of operational balloon missions used by NASA:  Conventional Ballooning, Long Duration Ballooning (LDB) , and Ultra Long Duration Ballooning (ULDB).  Conventional missions typically use direct line-of-sight  for command and data transmission and have flight durations ranging from a few hours to a few days.  A Long Duration Balloon mission normally traverses between continents or around the world for one circumnavigation, while using satellite-based  systems for command and data transmissions during flights lasting up to three weeks.  Ultra Long Duration Ballooning, enabled via Super-Pressure Balloons (SPB), has been designed to increase flight durations up to one hundred days at altitudes of $\sim30$ km.
This altitude is above $99.9$\% of the atmosphere.
This is the new frontier of scientific ballooning and will significantly increase the amount of data that can be collected in one balloon mission. NASA has already had several successful test SPB balloon flights, including one that lasted well over a month, and has begun to select science payloads for ULDB flights.

\subsection{Previous applications}
\label{sect:prevapp}


The range of applications of balloon-borne experiments and
instrumentation is extremely varied, as is their contribution to
scientific firsts.   Examples of science from balloon-borne
instruments include early maps of the anisotropies of the Cosmic
Microwave Background (CMB) \cite{cril}, the first identification of
antiprotons in cosmic rays \cite{maki}, early detection of gamma-ray
spectral lines from supernova 1987A \cite{rest}, the first
observation of positron emission lines from the galaxy \cite{john},
early detection of black-hole x-ray transients in the galactic center
region \cite{rick}, and observations of chlorofluorocarbons (CFCs)
and chlorine monoxide radicals in the stratosphere \cite{robi}.  More
recently, in November 2008, measurements of high-energy cosmic-ray
electrons by the Advanced Thin Ionization Calorimeter (ATIC)
instrument flown over Antarctica \cite{guzi} displayed a bump in the
spectrum that suggests a nearby cosmic-ray source, possibly the
signature of annihilation of dark-matter particles.  Many other
balloon instruments continue to produce significant discoveries.

In addition to the scientific measurements  enabled by balloons, there are many examples of spacecraft instrumentation that have been advanced by balloon-flight tests, through demonstrating detector technologies,providing complementary science data, and validating instrument concepts.
The design of both the Differential Microwave Radiometers (DMR; \cite{smoo})instrument on the COsmic Background Explorer (\emph{COBE}) satellite and the High Frequency Instrument (HFI; \cite{lama}) on the \emph{Planck} satellite benefitted from earlier balloon borne instruments (e.g, \cite{boug}; \cite{beno}) that returned science data and understanding of instrument performance. Other examples include the testing of Cd-Zn-Te (CZT) detectors in space-like environments, enabling their use on the highly-successful \emph{Swift} satellite \cite{romi}.

\begin{figure}[t]
\centering
\includegraphics[width=135mm]{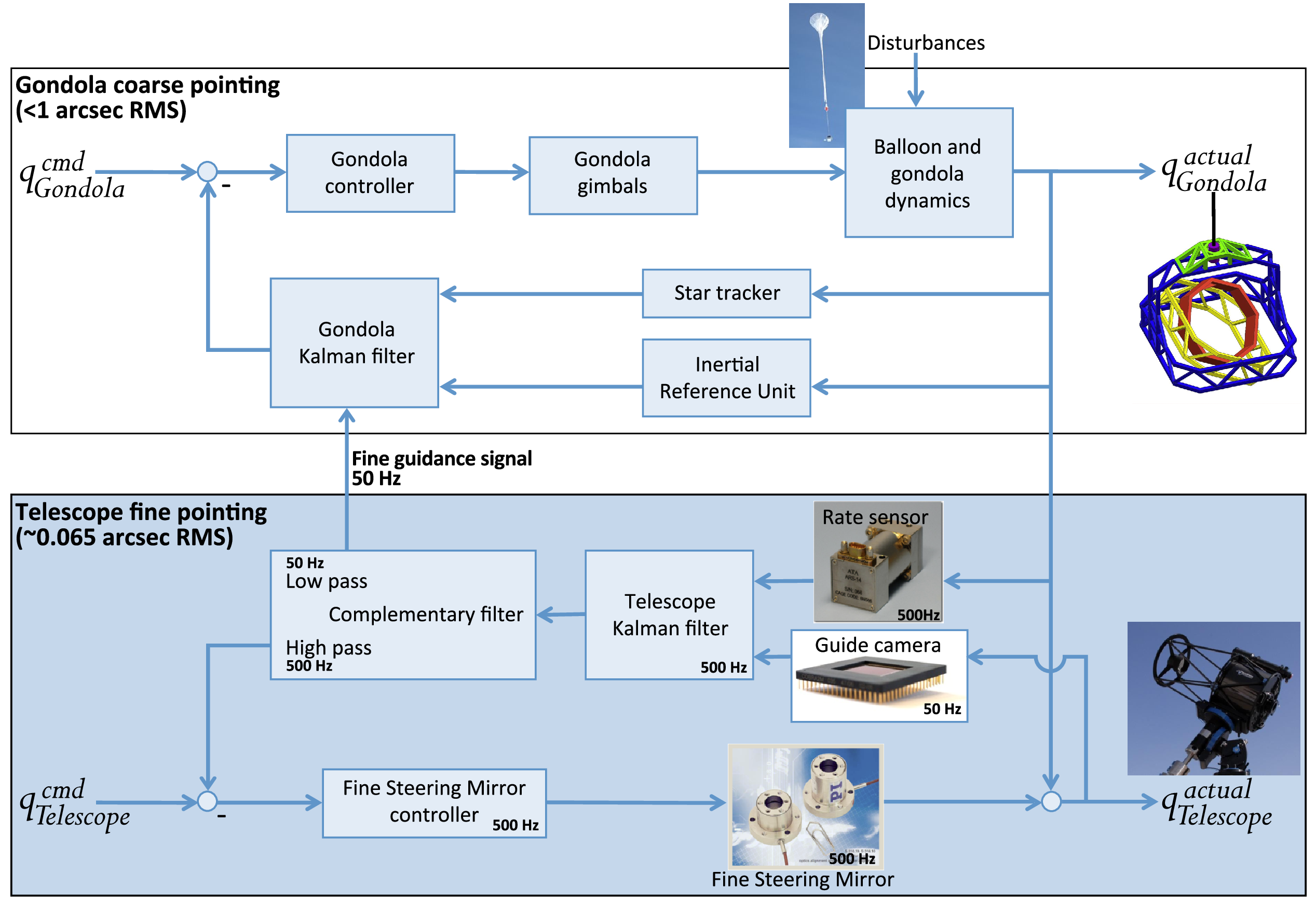}
\caption{A general multi-stage pointing control architecture for sub-arcsecond pointing stability. The top block refers to the gondola control  and the interface to the lower block, which is the instrument-level fine guidance system. The fine guidance system combines sensor information from guide CCDs and the angular rate sensors (ARSs) in order to drive a Fine Steering Mirror (FSM) control loop to correct for jitter.  This generates a fine guidance signal to the Wallops Arcsecond Pointing (WASP) system for improved telescope body pointing.}
\label{fig:flow}
\end{figure}

Planetary instrumentation has also been developed from balloon missions. The Mars Science Laboratory (MSL; e.g.\ \cite{umla}) will fly a Tunable Laser Spectrometer (TLS) that traces its heritage back to a series of balloon experiments that detected trace gases in the Earth's atmosphere using the Balloon-borne Laser In Situ Sensor (BLISS; \cite{menz}) instrument.  Indeed, the applications of balloon-borne experiments and instrumentation are as varied as their space-based counterparts.

\section{HALO requirements: \,\,\,\,  Attaining sub-arcsecond pointing and thermomechanical stability from balloon platforms}
\subsection{HALO Requirements}
\label{sect:tech}

While flights of balloon experiments and instrumentation have had great success, flight duration and robust pointing control have  limited their use for prolonged, high-precision optical astronomy.  Previous missions, such as the \emph{Sunrise} solar telescope \cite{bart}, employed the Sun as its tracking source and went on to demonstrate that pointing control with 0.05 arcsecond stability was possible.  Stratoscope II \cite{mcca} which began operations in the 1960's and achieved 0.02 arcsecond stability, showed that pointing control via star-tracking was possible using bright guide stars (5th - 9th magnitude).  Newer technological approaches, as discussed in this paper, together with the advent of ULDB flights hold the promise of bringing precision optical balloon astronomy to full fruition.

\begin{table}
\caption{The pointing stability and control requirements for HALO.  These are derived from the necessity for accurate shape measurements and PSF deconvolution for weak gravitational lensing. }
{\scriptsize
\begin{center}
    \begin{tabular}{ | p{6.4cm} | p{6.4cm} |}
    \hline
    Measurement requirement: & Instrument requirement:  \\ \hline
    0.06\arcsec 1 $\sigma$ RMS (0.15\arcsec FWHM) tip/tilt stability over 400 seconds with a goal of 0.04\arcsec RMS (0.1\arcsec FWHM)  & Centroid using guide stars down to AB magnitude 16 co-located in science Focal Plane Array (FPA) with frame transfer guide CCDs read at 50 Hz.  Fast Steering Mirror with 500Hz bandwidth. \\ \hline
    Correct for field rotation of sky and pendulation during 400 second integrations to 4'' & Combination of guide stars as above and  high rate inertial sensing (500Hz) on telescope structure with 3-axis angular rate sensors (ARS) sensitive to $<$ 50 nanoradians RMS \\ \hline
    For thermal/mechanical stability to PSF $\Delta$ellipticity $<$0.001, multiple measurements of guide stars over course of a night to characterize PSF stability & Stabilize telescope structure to ambient $\pm$4K to maintain stable alignment ($\pm$ 2K at optical instrument).  Also use  wavefront sensing software and actuated secondary mirror to maintain internal alignment.\\
    \hline
    \end{tabular}
\end{center}
}
\end{table}

Due to the nature of the proposed science, the HALO concept has stringent requirements on the pointing stability and control of the balloon platform. The primary measurement systematic in a weak lensing experiment arises from the need to accurately determine  galaxy shapes by
removing the distorting effects of the instrumental PSF from the measured shapes.
This drives requirements on the PSF size (\cite{paul}) and stability (\cite{amar}).  The specific pointing and thermo-mechanical stability requirements for HALO are displayed in Table 1. The first two requirements (on tip/tilt stability and rotational stability, or jitter) are derived from a desire to keep the PSF  close to diffraction limited in the optical, taking full advantage of the almost non-existent atmosphere (and thus seeing) at balloon altitude. In fact, the quantitative requirement is set so that the PSF size will have roughly equal contributions from diffraction, translation/rotational jitter, and charge diffusion within the CCD detectors.  Since these components are added in quadrature, minimizing any one of them has diminishing returns once they become roughly equal in size. The third requirement (on thermo-mechanical stability) is in place to ensure that the PSF stability (and thus knowledge) allows us to perform accurate shape measurements and PSF deconvolution.  The requirement is derived quantitatively from \cite{amar} so that the PSF measurement errors  in a hypothetical several thousand square degree HALO weak lensing survey would be sub-dominant to the statistical errors from such a survey.  Such a survey  could be carried out in one or two ULDB observing campaigns of $\sim100$ days each.  These requirements on PSF are designed to bring us close to achieving the PSF properties (size and stability) afforded by space without the full cost of a space mission.  Driving down the PSF size is important because PSF-systematics scale as the square of the PSF size (\cite{paul08}), affording HALO an order of magnitude advantage over ground-based surveys in this regard. Furthermore,  \cite{mass12} show that the measurement biases for galaxies shapes are a factor of several smaller for space-like PSFs than for ground-based PSFs, given the same shape measurement techniques, and \cite{amar} show that any even in optimistic observing conditions, future ground-based surveys will likely become systematics limited over 1000 square degrees.  Thus, the goal of this paper is to show that one aspect of a space mission (PSF size and stability) can be approached with a balloon-borne system.

\begin{figure*}[t]
\begin{center}$
 \begin{array}{cc}
\includegraphics[width=65mm]{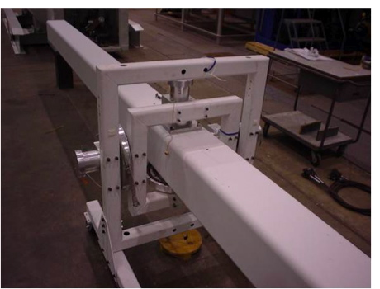}&
\includegraphics[width=66mm]{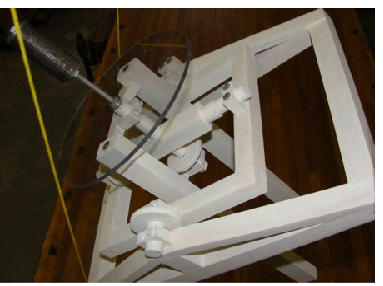}
\end{array}$
\end{center}
\caption{{\bf{Left:}} The WASP pointing system in a laboratory test with a mock telescope (long rod going from upper left to lower right) mounted in two gimballed frames (the white squares). This is the system used in the hang test described in the text with results shown in Figure~\ref{fig:WASP_results}.  The mock telescope is moved along the axes described by the silver metal bearings.    {\bf{Right:}}  Photograph of a scale model of the WASP system with the rotation axis added, allowing the telescope to correct for field rotation of the sky. The telescope can be cantilevered from a mount bearing in the center of the gimballed frames. The long white bar in the left panel has here been replaced with the metal rod extending from the middle of the picture to the upper left.  This rod represents a telescope and is on a rotation axis.}
\label{fig:WASP}
\end{figure*}

In this paper, we introduce a multi-stage pointing architecture that is capable of such sub-arcsecond pointing stability on balloon platforms.  This multi-stage architecture is depicted in Figure~\ref{fig:flow}.  We also outline an architecture that will meet the requirement on thermo-mechanical stability.   While the technical solution presented here is novel in its approach, its design is built upon the heritage of several existing mechanisms and balloon missions, and as such the implementation would not require exhaustive development.  We highlight that while this design solution was developed for the HALO concept, it has broader applications to any field of astronomy requiring high pointing stability (see section 4.2 for discussion).

\begin{figure*}[t]
\centering
\begin{center}$
 \begin{array}{cc}
\includegraphics{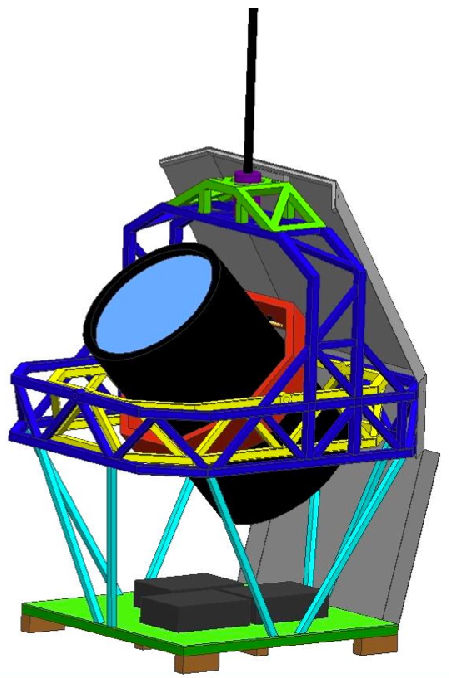}&
\includegraphics[width=85mm]{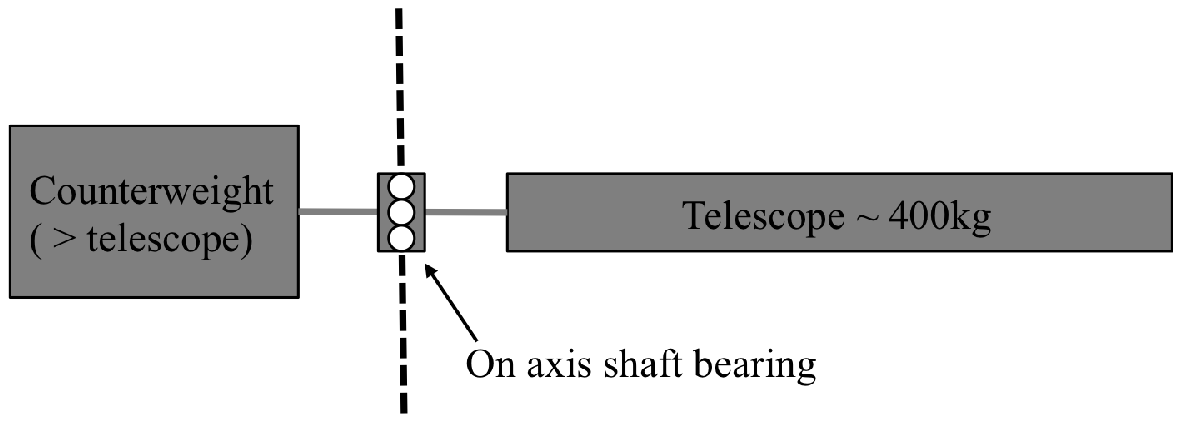}
\end{array}$
\end{center}
\caption{  Two different design approaches for the gondola structure.  In both cases the gondola and telescope design employs a nested series of gimballed frames in order to correct for jitter of the telescope.  Rotational correction (to counteract sky-rotation) is accomplished via either a  series of gimballed motions as described in Appendix A (left), or an on-axis shaft bearing with a corresponding telescope counterweight (right). The counterweight could consist of electronics, batteries and other components of HALO so as not to increase the total mass.}
\label{fig:HALO_gimbles}
\end{figure*}

\subsection{Nested-gimballed frames and gondola structure}
\label{sect:gondola}

We have considered several gondola structures that would provide the required range of motion for a wide field survey as required by the HALO science goals.
The viable design space 
separates into essentially two options.
A structure with 3 or 4 nested gimbals (left panel of Figure~\ref{fig:HALO_gimbles}; see also Appendix A)
can provide
azimuth, elevation, and roll control (the latter through
coordinated motion of two or more axes).
The second option (and the  HALO default design) is
the current Wallops ArcSecond Pointing (WASP, Section~\ref{sect:WASP}) design with a supplemental roll axis and counterweight as shown in the the right panels of Figures~\ref{fig:HALO_gimbles} and \ref{fig:WASP} (see also Figure~\ref{fig:WASP_flight}.
The gimbal designs (without counterweight) all require precision gondola azimuth axis control (the first gimbal). Control algorithms for the three-gimbal designs are somewhat complicated (again, see Appendix A), while in a four-axis system the gimbals move in a very simple way, but the gondola azimuth must be very accurately controlled. The required level of precision for this type of design has not yet been demonstrated. The current WASP design (without roll control) is straightforward and its performance demonstrated in  laboratory hang tests and a recent test flight with a dummy telescope. To control roll the WASP design must be augmented with an axial roll bearing with the instrument cantilevered out in front of the bearing and a counterweight behind the bearing. The rotational performance of such a system has not been demonstrated and there may be mass penalties due to the counterweight.
 The WASP team does have plans to  fully demonstrate this aspect of the design in the near future.
 The current HALO baseline is to use the WASP system with an axial roll bearing.  However, we are still considering three and four axis gimballed systems and describe the rotational mathematics of such systems in Appendix A.

\subsection{Gondola coarse azimuth control}

The first stage of a precision pointing architecture enables coarse attitude control of the gondola in which everything else is contained.
Systems such as Wallops Flight Facility's (WFF) advanced coarse rotator have been developed and are fully mature. The system has been used on multiple NASA balloon missions to position the entire balloon gondola for the purpose of orienting solar panels towards the sun and providing approximate positioning of the instrument payload. The accuracy of this rough pointing system is $\pm$ 5 deg in azimuth.
The software in the rotator has been recently enhanced to accept
azimuth data from a GPS attitude determination unit,
and can position the gondola to any
desired azimuth relative to the GPS attitude during the day or night to an accuracy within 1 deg.

\subsubsection{Wallops Arc-Second Pointing system (WASP)}
\label{sect:WASP}

\begin{figure*}[t]
\centering
\includegraphics[width=135mm]{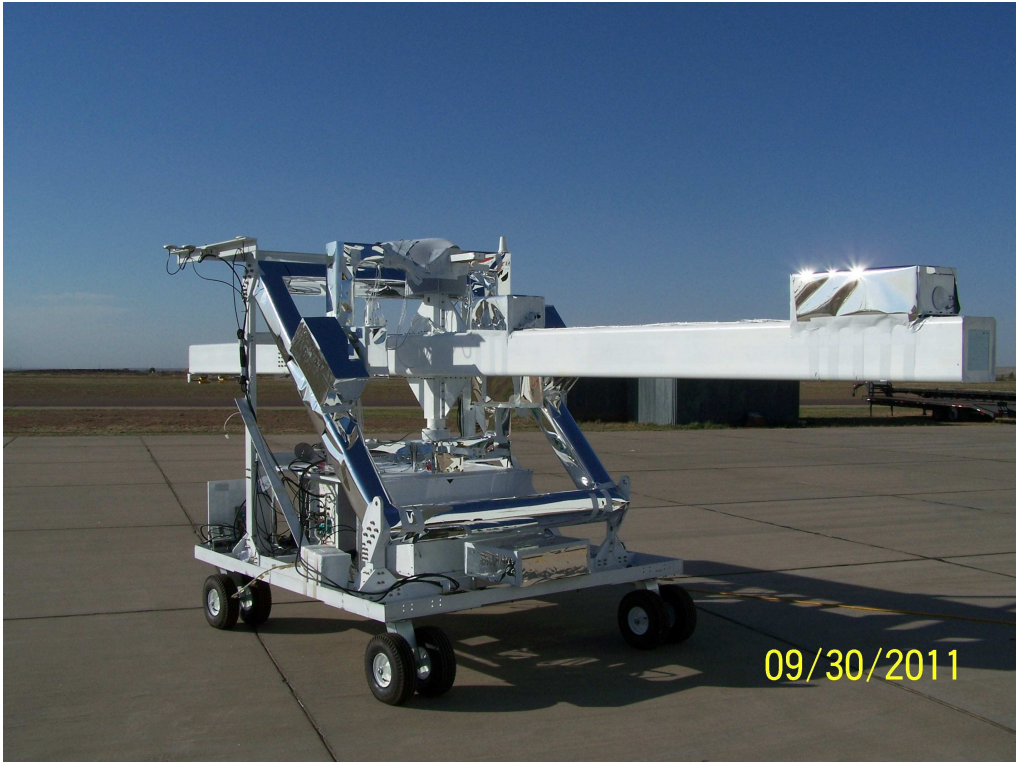}
\caption{The WASP system as flown on a test flight in the Fall of 2011.  This system showed pointing stability of $\sim0.3$\arcsec (RMS) in pitch and yaw using a mock telescope and full flight electronics. The system was operational for nearly 2 hours during a flight on an 11 million cubic foot balloon flying at $105,000$ feet. }
\label{fig:WASP_flight}
\end{figure*}

The baseline HALO design adopts the Wallops Arc-Second Pointing (WASP) system (which has 2 gimbals, see Figures~\ref{fig:WASP} and \ref{fig:WASP_flight}), augmented with an axial roll bearing.
The instrument will cantilevered out in front of the bearing and the counterweight could consist of batteries and electronics that are already part of HALO's mass budget.
The WASP team have plans to fully demonstrate this aspect of the design in the near future.

The WASP design approach recognizes that the elimination of static friction is essential, and that nonlinearities can cause serious errors in fine pointing. The design eliminates this static friction by keeping the bearing shaft constantly rotating using a small torque motor and planetary gear reduction box. The design approach also uses the balloon load train (the suspension cables and recovery parachute) as a reaction torque source.  This allows for continuous pointing by making use of the attachment of the gondola through the flight train to the balloon to absorb the momentum reacted from the telescope into the gondola.

The WASP stage is designed to utilize an attitude solution developed by
integrating the rates from an inertial rate sensor. During pointing
acquisition, the integrated solution is aided by a star-tracker which provides 3-axis attitude knowledge to the 3 arcsecond level. During
mission fine-pointing operations, the attitude solution is aided by
target offsets provided by the payload (at 50Hz).All stages of the pointing control have enough control authority for acquisition of the science target and guide stars  and hand-off to the fine guidance system for stabilization.


\begin{figure*}[t]
\centering
\includegraphics[width=140mm]{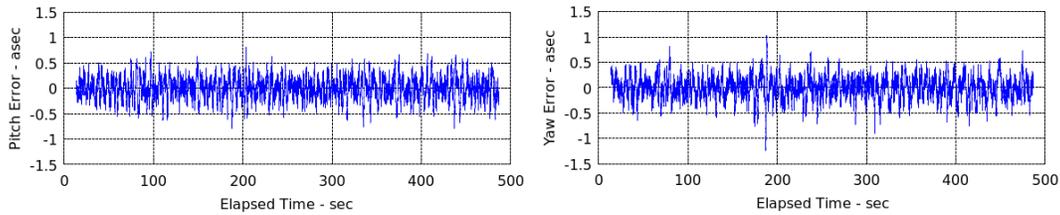}
\caption{Sample station-keeping data from a WASP hang test. This demonstrates the ability to achieve sub-arcsecond target error nulling (0.22\arcsec RMS) with the mechanical configuration displayed in Figure  \ref{fig:WASP} in a flight-like test environment.The flight-like test environment included hang tests where the WASP apparatus was freely swinging and added mechanical perturbations of the system like those that could be expected for a stratospheric balloon payload.}
\label{fig:WASP_results}
\end{figure*}

\subsubsection{WASP hang test and flight test results}
The system design consists of an orthogonal pair of pitch and yaw gimbals for control of the telescope structure. A prototype system was constructed, suspended, and performance tested at Wallops Flight Facility (see the left panel of Figure  \ref{fig:WASP}).  This consisted of a WASP Demonstration Prototype System with a simulated gondola/bridle system beneath a coarse azimuth pointer.
With this mechanical configuration, and in a flight-like test environment, it was demonstrated that a sub-arcsecond target error nulling of 0.22-asec (RMS) could be achieved (Figure~\ref{fig:WASP_results};  see also \cite{dewe}).  It was also demonstrated that the mutual interactions between the WASP and rotator control loops are negligible, i.e.\ the rotator is able to maneuver the gondola without significantly impacting performance of the WASP.  A full suite of tests was performed, including target slewing, station-keeping, step disturbance-torque response characterization, and large-angle disturbance inputs into the gondola (pendulous and wristing modes)\footnote{If we imagine
the flight train as an ``arm'' hanging below the balloon, the parachute
system is the upper-arm, the cable ladder is the fore-arm, the rotator
is the pivot point at the wrist, and the gondola is the hand. The
wristing mode consists of gondola rotations about the pivot at the
rotator (wrist). This mode is significant for the WASP.}.
Linear analysis,
nonlinear simulation and testing show that a significant vertical
imbalance in the instrument leads to an unstable rotational excitation
of the gondola below the rotator, since the corrective torques from the
control system will necessarily be in resonance with the natural
transverse frequencies of the flight train. Thus, stability of this mode
requires a combination of the slight natural damping of the flight train and tight specifications on vertical imbalance of the telescope.
Testing also included sequencing of mission modes from uncaging (unstowing the telescope), multiple target acquisitions and station-keeping, to recaging and shutdown.

A full engineering test flight was recently completed and showed gondola pointing stability of $\sim0.3$\arcsec (RMS) in pitch and yaw.  This is better than the HALO requirement, but the WASP test configuration contains a large inert mass (meant to simulate a telescope) which may help to damp vibrations. For this test, the  WASP system ground processor was replaced  with a single flight computer, and the commercial motor driver and resolver circuits were replaced with flight quality electronics.  All electronic systems worked as expected during the test flight.  The test flight, which took place on 7 October 2011, lasted $4.75$ hours and allowed the WASP system to be exercised for nearly 2 hours.  The flight was done using an 11 million cubic foot balloon at a float altitude of $105,000$ feet. Coarse rotation was performed using a rotator in solar tracking mode. As discussed in the following section, rotational control at the level required for HALO remains to be demonstrated. The WASP system used for the flight test is shown in Figure~\ref{fig:WASP_flight}.

\subsection{Telescope rotation}

WASP tests to date have focused on tip/tilt stability.  However, for a survey telescope like HALO, the stability of the resultant image data also depends upon correctly eliminating sky rotation,  the apparent effect of the sky rotating due to the Earth's rotation.  Hence, for even moderate length exposures, there is this additional pointing constraint and the system architecture must be designed to eliminate rotation about the telescope line-of-sight. To control image blurring at the level required for robust weak lensing measurements of HALO's target galaxy population, rotational jitter about the telescope line-of-sight must be less than one-fifth of a pixel at the edge of the half square degree HALO Field of View (FOV; 4~arcseconds FWHM).
This requirement on rotation is tighter than those on tip-tilt pointing because rotational jitter of the PSF is harder to model than translational modes; the effects of rotational motion on a star depend on the stars distance from the center of rotation, but translation motions  affect every star in the FOV equally.

In addition, the efficiency of survey applications like HALO is increased if the cameras do not rotate relative to the sky during an entire observation campaign. This ensures that subsequent exposures can be tessellated into a large mosaic.
For HALO, this also maintains the orientation of the Point Spread Function (PSF), which includes directional spikes from diffraction around secondary mirror supports. 
Even for observations of a single pointing, a system architecture capable of maintaining constant parallactic angle (the angle between a fixed direction on the telescope and the direction to the north celestial pole) will be able to track the same guide stars throughout the night.
This  guarantees high-performance centroiding and attitude control using standard control algorithms.

The default HALO design uses the WASP system plus an extra WASP bearing/motor placed at the rear of the telescope to directly rotate the telescope about the line of sight (see the right hand side of Figures~\ref{fig:HALO_gimbles} and \ref{fig:WASP}). However, as described in Section~\ref{sect:gondola}, we are considering other options for the rotational correction that do not include a motor at the rear of the telescope.  The assessment of the field rotation matrices and dynamics along with a comparison of 3- and 4-frame gondola designs is included as Appendix A.

\subsection{Fast Steering Mirror and star-tracking}

Thus far, we have discussed the first stage of the pointing-system, i.e. the gondola's coarse azimuth control, then the second stage WASP gondola-telescope three-axis gimballed pointing system, which provides azimuth, elevation, and roll control.
The third and final stage of our precision pointing architecture contains an angular rate sensor (ARS) on the telescope and guide CCDs in the focal plane. These sensors drive a Fast Steering Mirror (FSM) within the telescope, which provides line-of-sight tip and tilt control, and also feed back into the WASP gondola control system. The fine guidance stage stabilizes the line-of-sight tip and tilt angles to better than $\sim$ 0.06 arcseconds (RMS).  The complete multi-stage pointing system, showing the various feedback loops between the stages, is shown in Figure~\ref{fig:flow}.

HALO includes four frame-transfer guide CCDs located at the edges of the science focal plane.
The guide CCDs will track guide stars around the science observation field of interest. A star distribution study for probable HALO survey fields was carried out to identify the availability of suitable guide stars. This study found that with four guide CCDs, 97\% of pointings will have at least 2 guide stars of the required brightness ($8\leq$mag$\leq12.5$) within the guide CCD areas.
Two or more guide stars provide solutions for all 3 pointing errors of the line-of-sight (tip, tilt and roll). Furthermore, our assessment of the centroiding performance showed that the CCDs can be read at 50Hz while meeting the pixel centroiding performance dictated by the pointing stability requirements (Table 1).  For the small number of pointings without 2 sufficiently bright guide stars, a slower than nominal readout of the guide CCDs will provide sufficient S/N for guiding with an accuracy that degrades gracefully with decreasing readout frequency.

 The 3-axis ARS is mounted on the telescope barrel, takes angular rate measurements at 500Hz or more,  and  is used to measure disturbances from $>$10 Hz to 1 kHz.
The fine guidance CCDs track guide stars in the field and readout at $<$ 50 Hz to provide centroiding information at the science focal plane level.
The fine guidance CCDs sense the lower frequency ($<$10 Hz) jitter, calibrate the ARS bias and alignment errors, and provide an inertial attitude reference to point and roll the telescope. The ARS provides the high-rate relative attitude information needed for pointing stability. The measurements from the fine guidance CCDs and ARS are combined in a Kalman filter that estimates the line-of-sight pointing and rotation error. These errors are complementary filtered such that the mid-high frequency content (tip and tilt at $>$0.1Hz) is used to drive the FSM (at 50Hz closed loop bandwidth) and the low frequency (tip, tilt and roll at $<$0.1 Hz) content is sent as the fine guidance signal to drive the WASP system (5hz closed loop).

\begin{figure*}[t]
\centering
\begin{center}$
 \begin{array}{cc}
\includegraphics[width=68mm]{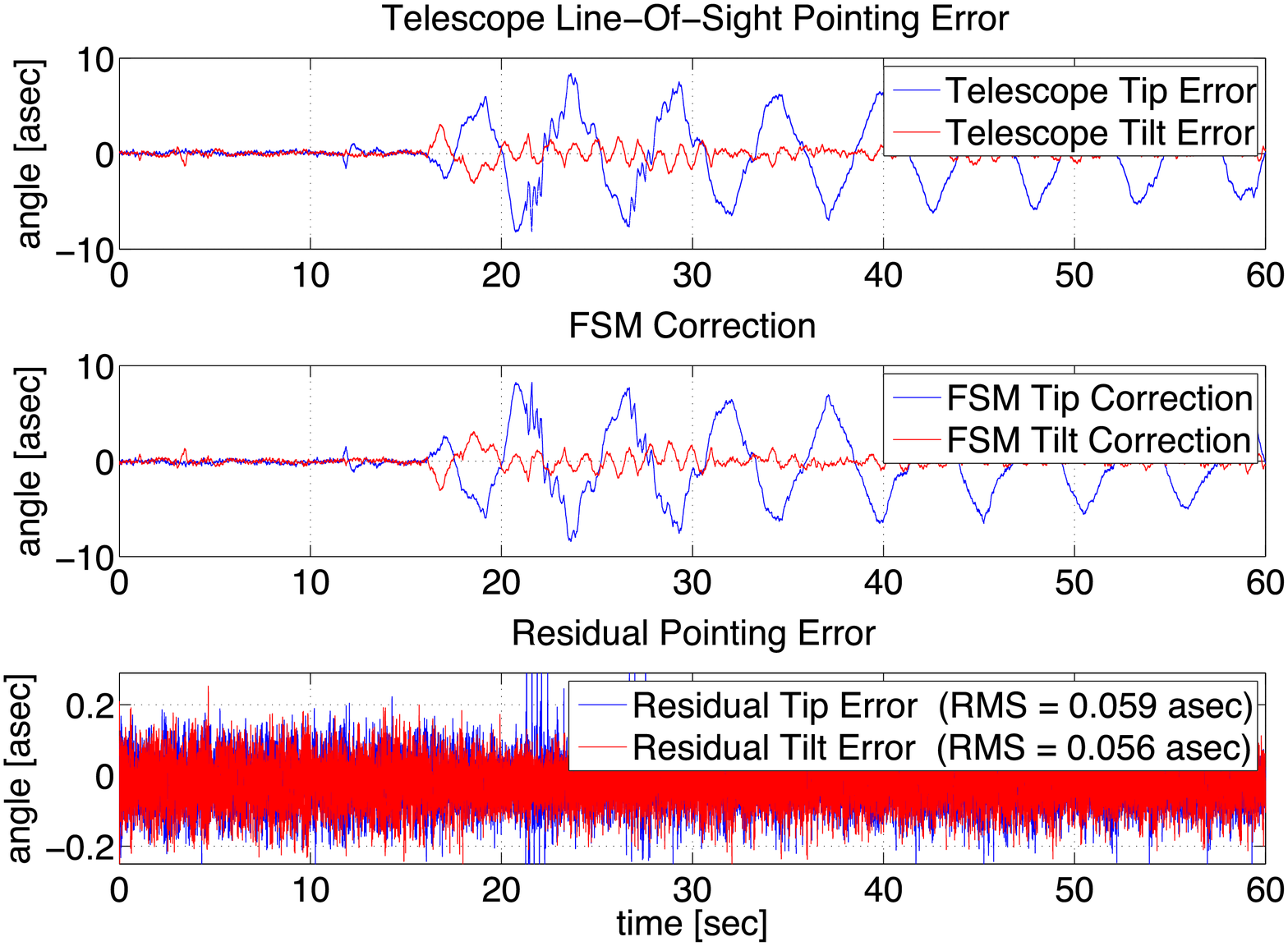}&
\includegraphics[width=68mm]{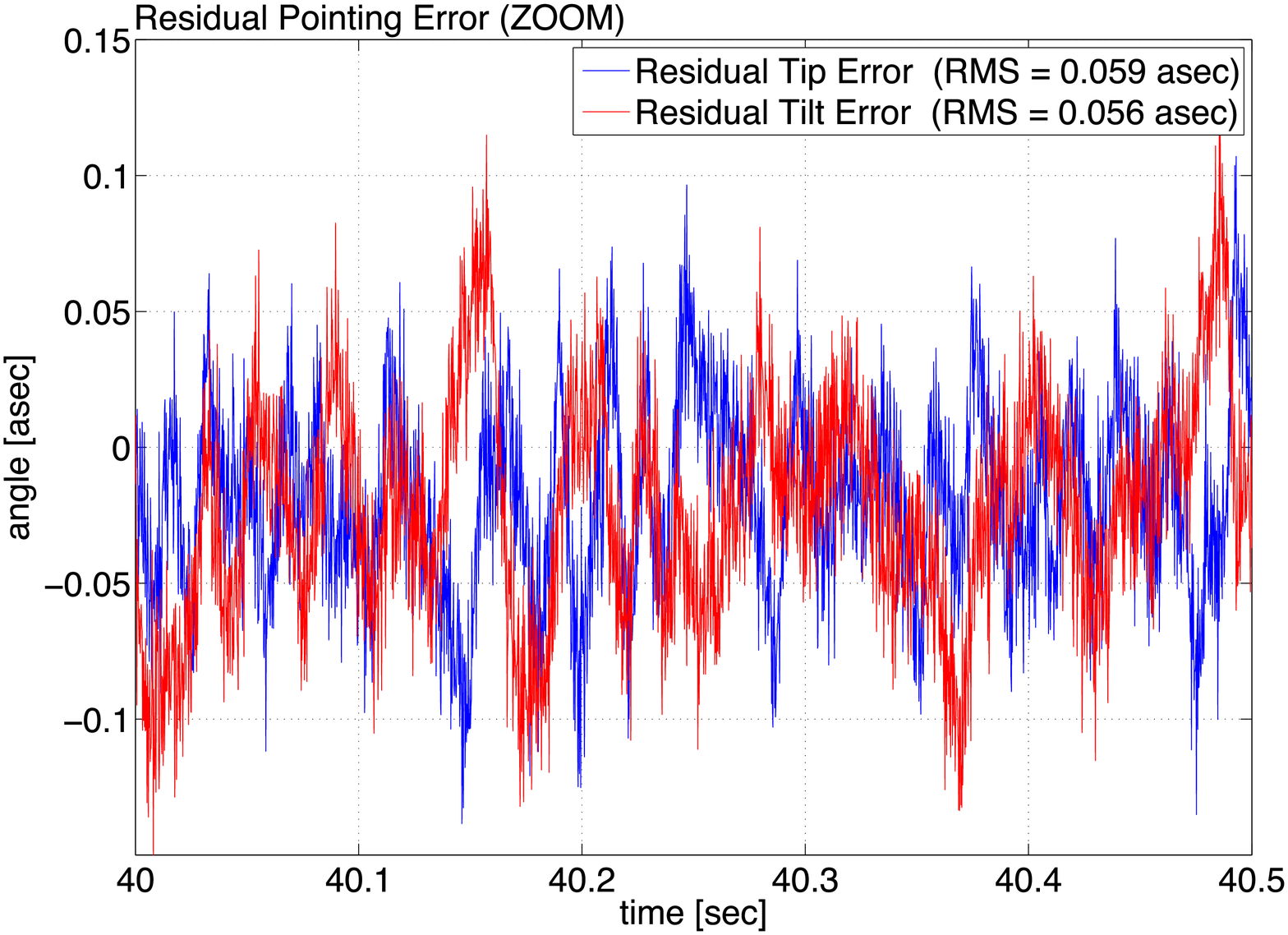}
\end{array}$
\end{center}
\caption{{\bf{Left:}}  Detailed simulations of the instrument-level fine guidance system show that the guide sensors and high-rate attitude sensors can combine to drive the FSM to correct for jitters down to below the $\sim$0.06 arcseconds RMS level. The top panel shows the experimentally-measured jitter signal of the WASP gondola-telescope after the first stage (gondola control) correction. These are inputs from the WASP hang test illustrated in Figure~\ref{fig:WASP_results}.  Taking this as our initial input, we show that we can make the necessary corrections with the FSM to achieve our required pointing accuracy. The central panel shows the simulated correction command to the FSM to suppress the jitter, and the bottom panel shows the residual error is significantly reduced.  {\bf{Right:}} A zoom of the residual high-frequency jitter from the bottom panel on the left shows more clearly that the RMS value achieved is $\sim$0.06 arcseconds (0.15 FWHM).}
\label{fig:FSM_sims}
\end{figure*}

\subsubsection{Fine Pointing Simulation}

A dynamic simulation model of the second-stage system has been developed to demonstrate that the tip-tilt jitter can be measured and corrected by the FSM to high precision ($\sim$0.060 arcseconds RMS -- see Figure ~\ref{fig:FSM_sims}).
The assumed disturbance model for our simulations was taken from experimental data from the WASP system (see Figure~\ref{fig:WASP_results}), which we used to generate raw line-of-sight pointing errors. These errors are consistent with those experienced by previous balloon missions \cite{pasc}.

The driving  signal sent to the FSM to correct for the input errors was modeled as a quaternion error signal (\mbox{i.e.} the tip/tilt error), which was converted into three actuator position commands.  This conversion represents the ``inverse kinematics solution" which was derived in closed-form using a linearized kinematic FSM model that accurately captures the kinematic degrees-of-freedom of the FSM.
The three actuator commands are passed through a (double-integrator, PII-type, 50Hz-bandwidth, 60$^{\circ}$-phase margin) ``jitter-reduction controller.''  A second integrator is added to this controller to effectively compensate for the losses introduced by the complementary filter.   The output signal from this filter block represents the three position commands that are sent to the FSM local controller hardware.    The FSM local controller  hardware is pre-programmed to process this input signal in either open-loop mode, or closed-loop mode.  For the open-loop case, the input represents the voltage applied to a Lead-Zirconate-Titanate (PZT) actuator to move the FSM.  In the closed-loop case, a local feedback sensor that directly measures the PZT actuator's stroke is used to create the error signal which is driven to zero at high-bandwidth/high-sampling (5kHz) by the FSM local closed-loop (Proportional-Integral) controller.  The  closed-loop local controller identified for HALO will be  used to attenuate the actuator hysteretic nonlinearity, which acts as a destabilizing internal disturbance for the actuator when it is driven in open-loop mode.  We modeled both the open-loop and closed-loop cases, both with and without a realistic actuator hysteresis model ($0.2\%$ nonlinearity).  Although the closed-loop approach was shown to reduce jitter substantially (by as much as 50\%, 1$\sigma$), we observe that the open-loop mode is sufficiently accurate to meet requirements.
In addition to verifying line of sight (LOS) accuracy, the simulation study also monitored the simulated performance of the FSM to confirm that it was conservatively within the specifications of commercial off the shelf (COTS) FSM hardware.   In particular, it was shown that only 2\% of the available actuator displacement capability was needed to perform the required LOS error attenuation.  Thus our preliminary assessment of the attitude estimation performance indicates that this approach will satisfy the HALO requirements (as is shown in Figure~\ref{fig:FSM_sims}).

\subsection{Structural Jitter}

Structural jitter induced by control structure interactions or balloon gondola dynamics could limit the pointing performance as was recently experienced by the \emph{Sunrise} mission \cite{bart10}, \cite{berk}. Their findings indicate the presence of an unmitigated 10Hz oscillation mode, plus many unmitigated high-frequency structural-jitter modes above 40Hz possibly caused by bearing rumble.
During HALO construction, we will perform modal tests and a closed-loop pointing test to measure structural vibrations.
If necessary we will add passive isolation to the telescope to gimbal the mounting interface, tuned mass dampers, and/or a passive damping implementation such as  properly tuned visco-elastic rotary motor mounts, which could help reduce structural jitter.  For example, visco-elastic rotary gimbal mounts at the interface between the inner-most gimbal and the telescope could be tuned to provide structural jitter attenuation
\cite{vonFlotow91}.

\subsection{Thermo-mechanical stability}

Maintaining a constant PSF to meet HALO's science goals will also require thermo-mechanical stability along the optical path.
Two effects dominate this budget.
First, the alignment of optical components can be affected by gravitational loading or temperature gradients. The shape of the PSF is most sensitive to differential motion of the telescope structure, for example expansion of one side of the telescope. To some degree, such motions can be compensated by the FSM and a movable secondary mirror; however, to reduce mechanical distortions to within an acceptable range (the third requirement in Table~1), the telescope structure and optical bench must be actively temperature-controlled.
Second, thermally induced airflow inside the optical path can cause ``mirror seeing''.
Preliminary balloon data from \cite{trau} indicate that  that turbulence induced by a warm primary mirror does not significantly degrade seeing in a high-altitude balloon environment. However, this data is not entirely conclusive, so we seek to implement a design that optimizes structural thermal stability and also minimizes heat gradients that might produce turbulent airflow.

HALO will launch at an ambient temperature of $\sim290$ K and rise to float altitude where the ambient temperature is $\sim240$ K and the pressure is a soft vacuum of a few torr. Solar radiative loading will vary by several hundred $\frac{W}{m^{2}} $ from daylight to night, while earth thermal irradiance varies with gondola orientation. In this environment, mechanical stability of the HALO telescope and instrument will be achieved through active temperature regulation of the structure and mirrors, aided by sun- and earth-shields which reduce the radiative loading to a small value. The structure will equilibrate to the ambient $240$ K temperature relatively quickly ($<12$ hours), despite being well-insulated. Thin-film heaters, on the structure under the insulation, will maintain the structure temperature stable to $\pm4$ K, slightly above the warmest ambient conditions, thereby maintaining mechanical alignment of the telescope structure. Maintaining the temperature slightly ``hot-biased'' is commonly done when the external temperature is expected to vary over a small range, or (particularly when using preset mechanical thermostatic switches) to account for uncertainty in foreknowledge of the ambient temperature. Rigid, lightweight insulation effective at the soft vacuum (few torr) ambient float pressure has been developed over the past decade under NASA sponsorship \cite{augu},
for cryogen and related storage and insulation applications; it is now a commercially-available product\footnote{Aspen Aerogels Cryogel x201, http://www.aerogel.com/products/pdf/Cryogel\_x201\_DS.pdf} with low mass and excellent insulating performance that will minimize both electrical demand for the heaters and heat loss to the surroundings.
The primary mirror will cool during ascent, but more slowly than the structure due to its mass. Approximately 30\% of the cooling will occur during the first hour of ascent, aided by forced convection over the back surface of the primary. By that time HALO will have approached $15$ km altitude, and the effectiveness of convective cooling will have decreased markedly. The remainder of the cooling, by a combination of radiation and natural convection (with reduced effectiveness but still present), will require an additional two days. Upon reaching final temperature, the primary will be temperature-controlled by electric heaters in a flexible sealed cavity on the mirror back surface. Cooling of the secondary mirror is treated similarly, but occurs more quickly because of its smaller size and more-exposed location. Preliminary pointing tests and calibration data can be taken during the first 2 days of a HALO flight, but full science operations will only begun once thermal equilibrium has been reached.

Maintaining the primary near ambient temperature has the added benefit of suppressing convection-induced seeing degradation. The primary front surface and optical path are enclosed by a lightweight, ``floating'' baffle which is sealed against air flow (by a highly flexible membrane) at the plane of the primary. This baffle shields the primary from external airflow and cools by radiation to ambient conditions, providing a stable air column within the telescope optical path. The rear surface of the primary and the instrument enclosure are separately temperature-controlled, again to near ambient temperature.

\section{Applications of stable balloon platforms}

There are a number of astronomy applications that require a pointing design that is stable at or near the level described here.

As discussed in the introduction, weak gravitational lensing is the systematic deflection of light from faint background galaxies by foreground dark matter structures. This distorts the apparent shapes of background galaxies by a few percent. This is too small to be measured in each galaxy but a mean signal can be obtained from similarly distorted galaxies along adjacent lines of sight.  Thus, weak lensing is a statistical measurement, requiring large wide-angle surveys to effectively constrain the properties of dark matter and dark energy (\cite{amar}; \cite{hu}; \cite{mkr}).  At the same time, as discussed in the introduction, due to the small galaxy shape distortions being measured, high image quality is essential. The large FOV and imaging precision attained by the HALO  architecture outlined here is 
designed to meet these requirements.

Searches for extra-solar planets also require an observing platform stable over long durations.
Planets are now known to exist outside our solar system in a variety of forms: including gas giants, hot-super-Earths in short period orbits, and ice giants.
The challenge for coming missions will be to find terrestrial planets, one half to twice the size of the Earth, in the habitable zone of their stars where liquid water and possibly life might exist (see e.g.\ \cite{boru}).
The \emph{transit} and \emph{microlensing} methods for exoplanet detection both require continuous, simultaneous monitoring of the brightnesses of hundreds of thousands of stars -- in a small number of pointings with a very large field of view.
Coronographic exoplanet studies require a smaller field of view, but still rely on the high stability enabled by the system described.
A balloon architecture close to HALO could perform these studies.

Galaxy surveys also have many compelling science goals that could take advantage of a stable balloon platform.
For example, the Cluster Lensing And Supernova survey with Hubble (CLASH; \cite{post}) aims to map the distribution of galaxies and dark matter in galaxy clusters, 
as well as to detect very distant supernovae (which must then be monitored from the ground due to the cost of space-based telescope time).
This survey will also allow study the morphological evolution and the internal structure of galaxies in and behind these clusters.
High resolution imaging is required to resolve these distant galaxies, and long duration pointing stability and control is needed to build up a statistically significant sample at a range of cosmological redshifts.
Such studies could be performed with a stable balloon telescope for far less cost than space-based platforms.

Finally, studies of stars within our galaxy and nearby galaxies suffer confusion due to seeing in ground based observations.
A stable space-based platform would open up a new, cheap method for long duration deep surveys of nearby stellar populations.

\section{Summary and discussion}

We have introduced a multi-stage, sub-arcsecond pointing architecture composed of coarse azimuth control of the gondola, a three-axis 
secondary pointing stage
with sky de-rotation, and a fine-guidance stage utilizing an angular rate sensor and fast steering mirror.  Long duration pointing stability of 0.15 arcseconds FWHM (0.06 RMS) should be attainable with this setup.  When combined with ultra long duration balloon flights via super-pressure balloons, our design has the potential to yield near space-quality data at a fraction of the the cost of a similar space-based facility.

One such balloon concept is the \emph{High Altitude Lensing Observatory}.  A weak gravitational lensing mission, HALO has the ultimate goal of surveying significant fractions of the night sky to high-precision in order to constrain the nature of Dark Matter and Dark Energy.  The experiment design outlined in this paper will allow HALO to reach this ambitious goal.
The reusable telescope will be able to perform surveys of 1000--2000 square degrees per year at $\sim$1--2\% the cost of 
a dedicated space-based survey mission.

\section*{Acknowledgments}

The authors wish to thank to Joan Ervin, Roger Lee, Tanaz Mozafari, Barth Netterfield, Wes Traub, Chris Stoughton and David Pierce for work and discussions relating to the HALO concept.  RJM acknowledges financial support through STFC Advanced Fellowship PP/E006450/1.  The work of JR, BMD, JB, KL, PB, EJ, JA and CP was carried out at the Jet Propulsion Laboratory, California Institute of Technology, under contract with the National Aeronautics and Space Administration (NASA).  EJ acknowledges the support of ORAU under contract with NASA.

\section*{Appendix A: HALO Field Rotation and Gimbal Designs}

The design space we are considering for HALO separates into  into three- or four-gimbal designs or the WASP system with a supplemental roll axis and counterweight. We present here the mathematics behind the rotational transformations needed to achieve roll control for a telescope like HALO using three and four-gimbal designs.

In order to determine the required motion about  physical rotation axes, we equate the rotation matrix describing motion about these axes to the one describing motion of the target. From this comparison, it is possible to analyze angular position of each axis frame as a function of location of the target along its path through the sky, as parameterized by declination (dec) and Hour Angle (HA). The goal is to maintain constant orientation of the image (constant direction to the north celestial pole). This is achieved by converting a rotation of the telescope about the north celestial pole to a rotation about the physical gondola axes.

Rotation of $\theta$ counterclockwise about a unit vector axis $v$ $=$ $[a~b~c]$  is given by a rotation matrix
\begin{equation}
R_{v}(\theta) =  Q \sin(\theta) + (I - P)\cos(\theta) + P
\end{equation}
where
\begin{equation}
Q = \left(\begin{array}{ccc}0 & c & -b \\-c & 0 & a \\b & -a & 0\end{array}\right)
\end{equation}
with $P$ =  $v$$^{T}$$v$, and $I$ =  3 $\times$ 3 identity matrix.  A vector $p$ under a rotation described by $R$ is given by
\begin{equation}
p_{\mathrm{rot}} =  p_{\mathrm{original}}R~.
\end{equation}
Sequential rotations can be expressed as a product of the rotation matrices that compose it. It is important to be careful that the sub-rotations are ordered in the right way and expressed with respect to the correct axes, because rotation matrices are not multiplicatively commutative. Sequential rotations can be described in two ways. The next rotation is added intrinsically (with respect to moving frame) as a left product about intrinsic axes, and extrinsically (with respect to stationary frame) as a right product about extrinsic axes.

For example, extrinsic rotation about the x-axis [1 0 0] by 90$^{o}$ followed by rotation about z-axis [0 0 1] by -90$^{o}$ is described by
\begin{equation}
R =  R_{x}(90^{o})R_{z}(-90^{o}),
\end{equation}
which is equivalent intrinsically to a rotation of -90$^{o}$ about z followed by a rotation of 90$^{o}$ about the once-rotated x-axis, actually now at
\begin{equation}
[1,0,0] R_{z}(-90^{o}) =  [0,-1,0].
\end{equation}

Rotation matrices are typically named according to their intrinsic composition. This would be a ZX rotation.  We define a fixed coordinate system based on zenith as $z$, terrestrial north as $y$, and terrestrial east as $x$.  First we describe rotation of the sky. Stable sky pointing motion is described by
\begin{equation}
R_{\mathrm{pointing}} =  R_{x}(-\mathrm{dec})R_{y}(-\mathrm{HA})R_{x}(\mathrm{lat})
\end{equation}
where HA has been converted from units of sidereal hours to units of degrees (24 sidereal hours = 360$^{\rm o}$). This describes a rotation about $x$ (= East) of lat$^{\rm o}$, then about once-rotated $y$ (= North Celestial Pole) of -HA$^{\rm o}$, and finally about twice-rotated $x$ (= away from celestial north) of -dec$^{\rm o}$ in order to get to a target currently located at some declination and HA. As time passes, the target HA increases.
	
Sky motion with hanging disturbances superimposed is described by
\begin{equation}
R_{\mathrm{distpoint}} =  R_{\mathrm{disturb}} R_{\mathrm{pointing}}
\end{equation}
where
\begin{equation}
R_{\mathrm{disturb}} =  R_{z}(\theta_{\mathrm{roll}})R_{x}(\theta_{\mathrm{pitch}})R_{y}(\theta_{\mathrm{yaw}}),
\end{equation}
and roll, pitch, and yaw intrinsically define the disturbance motion of the telescope. The fine guidance CCDs on HALO will sense this disturbance as a shifting of the image, which can then be translated to roll, pitch, and yaw of the telescope.

\begin{figure*}[t]

\begin{center}$
 \begin{array}{ccc}
\includegraphics[width=44mm]{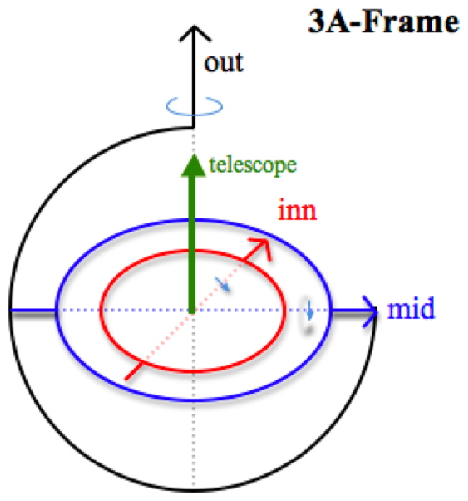}&
\includegraphics[width=44mm]{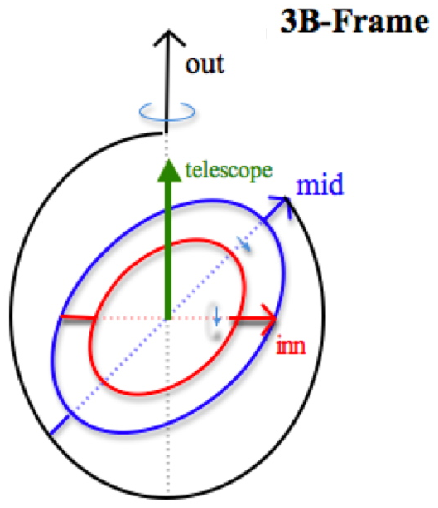}&
\includegraphics[width=44mm]{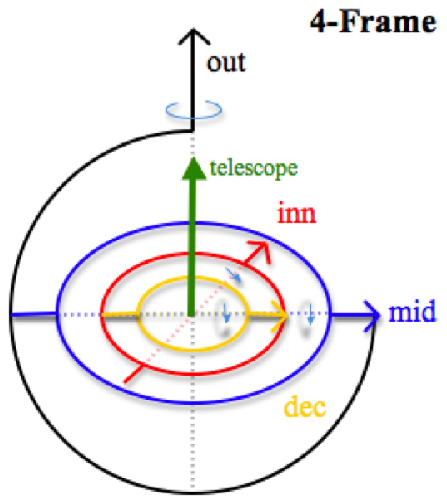}
\end{array}$
\end{center}

\caption{{\bf Left:} 3-frame design in orientation A:  ZXY configuration. Initial position with frame rotation angles = 0 when inner axis along $y$ (red), mid along $x$ (blue), outer along $z$ (black), and telescope also up (green).  {\bf Middle:} 3-frame design in orientation B:  ZYX configuration. Initial position with frame rotation angles = 0 when inner axis along $x$ (red), mid along $y$ (blue), outer along $z$ (black), and telescope also up (green):  equivalent to 3A rotated about $z$ by 90$^{\rm o}$ and with inner axis inverted.  {\bf Right:} 4-frame design. Initial position with frame rotation angles = 0 when declination axis along $x$ (yellow), inner along $y$ (red), mid along $x$ (blue), outer along $z$ (black), and telescope also up (green):  similar to 3A with an additional inner frame.}
\label{fig:gimbal_designs}
\end{figure*}

Next we define rotation of the gimbal frames. This consists of a product of rotations about gimbal axes:  described extrinsically inner to outer, which is equivalent intrinsically to outer to inner
\begin{equation}
R_{\mathrm{frames}} =  R_{\mathrm{innermost}} \, \times R_{\mathrm{outermost}}~.
\end{equation}
The rotation axes are defined in their initial positions with respect to the reference frame. The scenarios that follow are set up with each of these unit vectors along the $x$-, $y$-, or $z$-axis. This makes the product of matrices relatively simple such that the frame angles can be solved for explicitly by examining certain entries in $R_{\mathrm{frames}}$.

Any rotation matrices that describe the same ultimate rotation motion must be equivalent, though they may get there along different axes. In order to transform target motion into motion of gimbals, the rotation matrices for sky and frames are equated. By solving this system of equations we derive frame rotation angles as targets are tracked across the sky, and from these quantities also estimate rates of frames' rotations
\begin{equation}
R_{\mathrm{frames}} =  R_{\mathrm{sky}}~.
\label{equ:frames}
\end{equation}
With this it is possible to examine steady sky rotation to get a feel for target tracking, and examine sky rotation with roll disturbance to assess the effect of LOS rotation on frame angles.

Next, we examine three gimbal designs consisting of nested frames. Diagrams of their setup are displayed in Figure \ref{fig:gimbal_designs} and their frame rotation matrices are described below. The frame axes' initial alignment with the $x$-, $y$-, or $z$-axes results in a relatively simple rotation matrix that allows the frame angles to be solved for explicitly from certain entries $R_{row}$ column.

For the 3-frame gimbal system (Figure \ref{fig:gimbal_designs}: left), the rotation matrix describing frame motion is
\begin{equation}
R_{\mathrm{3frames}} =  R_{y}(\theta_{i}) R_{x}(\theta_{m}) R_{z}(\theta_{o}).
\end{equation}
This is equivalent to rotation of $\theta_{\rm o}$ about the outer frame axis, followed by a rotation of $\theta_{m}$ about the (once rotated) middle frame axis, followed by a rotation of $\theta_{i}$ about the (twice rotated) inner frame axis. The axes of motion are defined with respect to their initial position to the reference frame, which puts the outer, mid, and inner axes along $z$, $x$, and $y$ respectively, resulting in a ZXY rotation. Frame angles are solved for explicitly as
\begin{eqnarray}
\theta_{o} &=& \mathrm{arctan}(-R21 / R22)\\
\theta_{m} &=& \mathrm{arcsin}(R23)\\
\theta_{i} &=& \mathrm{arctan}(-R13 / R33)~.
\end{eqnarray}

For the 3B gimbal system (Figure \ref{fig:gimbal_designs}: middle), the rotation matrix describing frame motion is
\begin{equation}
R_{\mathrm{3Bframes}} =  R_{x}(\theta_{i}) R_{y}(\theta_{m}) R_{z}(\theta_{o})
\end{equation}
This is still equivalent to rotation of $\theta_{\rm o}$ about the outer frame axis, followed by a rotation of $\theta_{m}$ about the (once rotated) middle frame axis, followed by a rotation of $\theta_{i}$ about the (twice rotated) inner frame axis, though the axes positions have changed. The outer, mid, and inner axes are along $z$, $y$, and $x$ respectively, resulting in a ZYX rotation. The frame angles are solved for explicitly as
\begin{eqnarray}
\theta_{o} &=& \mathrm{arctan}(R12 / R11)\\
\theta_{m} &=& \mathrm{arcsin}(-R13)\\
\theta_{i} &=& \mathrm{arctan}(R23 / R33)~.
\end{eqnarray}

For the 4-frame gimbal system (Figure \ref{fig:gimbal_designs}: right), the rotation matrix describing frame motion is
\begin{equation}
R_{\mathrm{4frames}} =  R_{x}(\theta_{\mathrm{dec}}) R_{y}(\theta_{i}) R_{x}(\theta_{m}) R_{z}(\theta_{o})~.
\end{equation}
This is equivalent to rotation of $\theta_{\rm o}$ about the outer frame axis, followed by a rotation of $\theta_{m}$ about the (once rotated) middle frame axis, followed by a rotation of $\theta_{i}$ about the (twice rotated) inner frame axis, and finally a rotation of $\theta_{i}$ about the (thrice rotated) declination frame axis. The outer, mid, inner, and dec axes are along $z$, $x$, $y$, and $x$ respectively, resulting in a ZXYX rotation. However, it is slightly simpler than this. The declination frame is only used for declination selection, and not driven otherwise, so we set
$\theta_{\mathrm{dec}}=-\mathrm{dec}^{\circ}$. Then left-dividing both sides of the equivalence equation (\ref{equ:frames}) by $R_{x}(-\mathrm{dec})$ makes the outer, mid, and inner gimbals of the 4-Frame behave as the 3A-Frame system. We refer to this altered rotation as $R^{*} = R_{x}(-\mathrm{dec})/R$. The frame angles then are solved for explicitly as
\begin{eqnarray}
\theta_{o} &=& \mathrm{arctan}(-R^{*}21 / R^{*}22)\\
\theta_{m} &=& \mathrm{arcsin}(R^{*}23)\\
\theta_{i} &=& \mathrm{arctan}(-R^{*}13 / R^{*}33)\\
\theta_{\mathrm{dec}} &=&  - \mathrm{dec}~.
\end{eqnarray}
This 4-Frame design is essentially an equatorial telescope mount. The inner axis acts as right ascension, kept pointing at the celestial pole by the outer and mid frames. Only this inner right ascension frame needs to be driven to track the target and the declination axis is only changed to select the target.



\begin{thebibliography}{00}
\bibitem[Albrecht et al., 2006]{albr} Albrecht et al., 2006, ``Report of the Dark Energy Task Force'', arXiv:astro-ph/0609591
\bibitem[\protect\citeauthoryear{Peacock et al.}{2006}]{peac} Peacock, J., et al. 2006, ``Report by the ESA-ESO Working Group on Fundamental Cosmology'', astro-ph/0610906
\bibitem[\protect\citeauthoryear{Fu et al.}{2008}]{fu} Fu, L., et al. 2008, A\&A, 479, 9
\bibitem[\protect\citeauthoryear{Tereno et al.}{2009}]{tere} Tereno, I., et al. 2009, A\&A, 500, 657
\bibitem[\protect\citeauthoryear{Heymans et al.}{2012}]{heym} Heymans, C., et al. 2012, MNRAS, 421, 381
\bibitem[\protect\citeauthoryear{Scoville et al.}{2000}]{scov} Scoville et al., 2007, ApJS 172, 150S
\bibitem[\protect\citeauthoryear{York et al.}{2000}]{york} York, D. G., et al. 2000, AJ, 120, 1579
\bibitem[\protect\citeauthoryear{Crill et al.}{2003}]{cril} Crill B., et al., 2003, ApJS, 148, 527
\bibitem[\protect\citeauthoryear{Makida et al.}{1995}]{maki} Makida, Y., et al, 1995, IEEE Trans. Applied Superconductivity, 5, 658
\bibitem[\protect\citeauthoryear{Rester et al.}{1989}]{rest} Rester, A.C., et al, 1989, ApJ, 342, 71
\bibitem[\protect\citeauthoryear{Johnson, Harnden, \& Haymes}{1972}]{john} Johnson W. N., III, Harnden, F. R., \& Haymes R.C., 1972, ApJ, 172, L1
\bibitem[\protect\citeauthoryear{Ricker et al.}{1976}]{rick} Ricker, G. R., et al, 1976, ApJ, 204, 73
\bibitem[\protect\citeauthoryear{Robinson et al.}{2004}]{robi} Robinson, A. D., et al, 2004, Atmos. Chem. Phys. Discuss., 4, 7089
\bibitem[\protect\citeauthoryear{Guzik}{2008}]{guzi} Guzik, T. G., 2008, COSP, 37, 1138
\bibitem[\protect\citeauthoryear{Smoot et al.}{1990}]{smoo} Smoot, G., et al., 1990, ApJ, 360, 685
\bibitem[\protect\citeauthoryear{Lamarre et al.}{2010}]{lama} Lamarre, J-M., et al., 2010, A\&A, 520, 9
\bibitem[\protect\citeauthoryear{Boughn et al.}{1990}]{boug} Boughn, S.,  et al., 1990, RScI, 61,158
\bibitem[\protect\citeauthoryear{Beno\^{i}t et al.}{2002}]{beno} Beno\^{i}t, A., et al., 2002, APh, 17, 101
\bibitem[\protect\citeauthoryear{Roming et al.}{2005}]{romi} Roming, P. W., et al., 2005, SSR, 120, 95
\bibitem[\protect\citeauthoryear{Umland et al.}{2004}]{umla} Umland, J. W., et al., 2004, IEEE Aerospace Conference, March 2004.
\bibitem[\protect\citeauthoryear{Menzies et al.}{1983}]{menz} Menzies, R. T., 1983, Applied Optics, 22, 17, 2655
\bibitem[\protect\citeauthoryear{Barthol et al.}{2008}]{bart} Barthol, et al. 2008, Adv. Space Res., 42, 70-77
\bibitem[\protect\citeauthoryear{McCarthy}{1969}]{mcca} McCarthy, D. 1969, ``Operating Characteristics of the Stratoscope II Balloon-Borne Telescope'', IEEE Transactions on Aerospace and Electronic Systems, AES-5, issue 2, 323-329
\bibitem[\protect\citeauthoryear{Paulin-Henriksson et al.}{2009}]{paul} Paulin-Henriksson, S., et al., 2009, A\&A, 500, 647
\bibitem[\protect\citeauthoryear{Amara \& Refregier}{2007}]{amar} Amara, A., \& Refregier, A., 2007, MNRAS, 381, 1018
\bibitem[\protect\citeauthoryear{Paulin-Henriksson et al.}{2008}]{paul08} Paulin-Henriksson, S., et al., 2008, A\&A, 484, 67
\bibitem[\protect\citeauthoryear{Massey et al.}{2012}]{mass12} Massey., R, et al., 2012, in prep for MNRAS
\bibitem[\protect\citeauthoryear{DeWeese \& Ward}{2006}]{dewe} DeWeese, K., and Ward, P., 2006, 36th COSPAR Scientific Assembly, COSPAR, 36,2531
\bibitem[\protect\citeauthoryear{Barthol et al.}{2010}]{bart10} Barthol et. al., 2010, ``The Sunrise Mission", Solar Physics.
\bibitem[\protect\citeauthoryear{Berkfeld et al.}{2010}]{berk} Berkfeld et al., 2010, ``The Wave-Front Correction System for the Sunrise Balloon-Bourne Solar Observatory", Solar Physics.
\bibitem[\protect\citeauthoryear{von Flotow et al.}{1991}]{vonFlotow91}  von Flotow, A.H., et al, `` A Case Study in Passive Piezoceramic, Viscous and Viscoelastic Damping," Proceedings of the International Symposium on Active Materials and Adaptive Structures, Alexandra, VA, Nov. $4-8$, 1991
\bibitem[\protect\citeauthoryear{Pascale}{2008}]{pasc} Pascale, et al., 2008, ApJ, 681, 400
\bibitem[\protect\citeauthoryear{Traub \& Chen}{2007}]{trau} Traub \& Chen, 2007, ``Planetscope Precursor Experiment'', AAS Meeting 211, 30.06, Bulletin of the American Astronomical Society, 39:782
\bibitem[\protect\citeauthoryear{Augustynowicz \& Fesmire}{2000}]{augu} Augustynowicz, S., \& Fesmire, J., 2000, Advances in Cryogenic Engineering, vol 45, pp 1691-1698
\bibitem[\protect\citeauthoryear{Hu}{1999}]{hu} Hu, W., 1999, ApJ, 522L, 21
\bibitem[\protect\citeauthoryear{Massey, Kitching, \& Richard}{2010}]{mkr} Massey, R., Kitching, T., \& Richard, J., 2010, RPPh, 73,6901
\bibitem[\protect\citeauthoryear{Borucki et al.}{2009}]{boru} Borucki W., et al., 2009, IAUS, 253, 289
\bibitem[\protect\citeauthoryear{Postman et al.}{2011}]{post} Postman M., et al., 2011, arXiv:1106.3328


\end{thebibliography}
\end{document}